\documentclass{aa}

\newcommand{\grp}    {${\rlap.}^{\circ}$}
\newcommand{\pri}    {${\rlap.}^{\prime \prime}$}
\newcommand{\rl}     {${\rlap.}^{s}$}

\newcommand{\ltsima} {$\; \buildrel < \over \sim \;$}
\newcommand{\simlt}  {\lower.5ex\hbox{\ltsima}}            
\newcommand{\gtsima} {$\; \buildrel > \over \sim \;$}
\newcommand{\simgt}  {\lower.5ex\hbox{\gtsima}}            

\newcommand{\tyc} {TYC 3683-985-1}
\newcommand{\fgl} {3FGL J0133.3+5930}
\newcommand{\sw} {SWIFT J0132.9+5932}
\newcommand{\rosat} {1RXS J013326.9+592946 }
\newcommand{\twomass}{2MASS 01325529+5932158}
\newcommand{\nvss}{NVSS J013255+593217}

\usepackage{graphicx}
\usepackage{txfonts}
\begin{document}

\title{Peculiar objects towards \fgl: \\   an eclipsing Be star and  an active galactic nucleus}

\titlerunning{Peculiar objects towards \fgl}

   \authorrunning{Mart\'{\i} et al.}
 

   \author{
        Josep Mart\'{\i}\inst{1}
          \and     
          Pedro L. Luque-Escamilla\inst{2}  
          \and
          Josep M. Paredes\inst{3}
          \and
          Kazushi Iwasawa\inst{4,5}
          \and
          Daniel Galindo\inst{3}
          \and
          Marc Rib\'o\inst{3}
          \and
          V\'{\i}ctor Mar\'{\i}n-Felip\inst{1}
           }
          
\institute{
Departamento de F\'{\i}sica, Escuela Polit\'ecnica Superior de Ja\'en, Universidad de Ja\'en, Campus Las Lagunillas s/n, A3, 23071 Ja\'en, Spain\\
   \email{jmarti@ujaen.es, vmarin@ujaen.es}
   \and
Departamento de Ingenier\'{\i}a Mec\'anica y Minera, Escuela Polit\'ecnica Superior de Ja\'en, Universidad de Ja\'en, Campus Las Lagunillas s/n, A3, 23071 Ja\'en, Spain\\
  \email{peter@ujaen.es}   
  \and
  Departament de F\'{\i}sica Qu\`antica i Astrof\'{\i}sica,  Institut de Ci\`encies del Cosmos, Universitat de Barcelona, IEEC-UB, Mart\'{\i} i Franqu\`es 1, E-08028 Barcelona, Spain\\
  \email{jmparedes@ub.edu,  dgalindo@am.ub.es,  mribo@ub.edu}
  \and
Institut de Ci\`encies del Cosmos (ICCUB), Universitat de Barcelona (IEEC-UB), Mart\'i i Franqu\`es, 1, 08028 Barcelona, Spain\\
\email{kazushi.iwasawa@icc.ub.edu}
\and
ICREA, Pg. Llu\'is Companys 23, 08010 Barcelona, Spain\\
 }

   \date{Received XXXXXXX XX, 2016; accepted XXXXXXX XX, 201X}

 
  \abstract
   {}
   {We aim to contribute to the identification of unassociated gamma-ray sources in the galactic plane to enlarge the currently known
   population of gamma-ray binaries and related systems, such as radio-emitting X-ray binaries and microquasars. These objects are currently regarded
   as excellent test beds for the understanding of high-energy phenomena in stellar systems.}
   {Potential targets of study are selected based on  cross-identification of the third {\it Fermi} Large Area Telescope catalogue with 
   historical catalogues of luminous stars that have often been found as optical counterparts in known cases. Follow-up observations and analysis
   of multi-wavelength archival data are later used to seek further proofs of association beyond the simple positional agreement.} 
   {Current results enable us to present here the case of the {\it Fermi} source \fgl\ where two peculiar objects have been discovered inside its region of
   uncertainty.  One of them is the star \tyc\ (LS I +59~79) whose eclipsing binary nature
    is reported in this work.
   The other is the X-ray source \sw, which we found to be a likely low-power  active galactic nucleus  at $z=0.1143 \pm 0.0002$.
   If this second object is of blazar type, it could easily account for the observed gamma-ray photon flux.
   However, this is not confirmed at present, thus
   rendering the star system \tyc\ as a still possible alternative counterpart candidate to the {\it Fermi} source.
      }
   {}
  
   
   \maketitle
%

\section{Introduction}

\begin{figure*}
   \centering
   \includegraphics[angle=0,width=18.75cm]{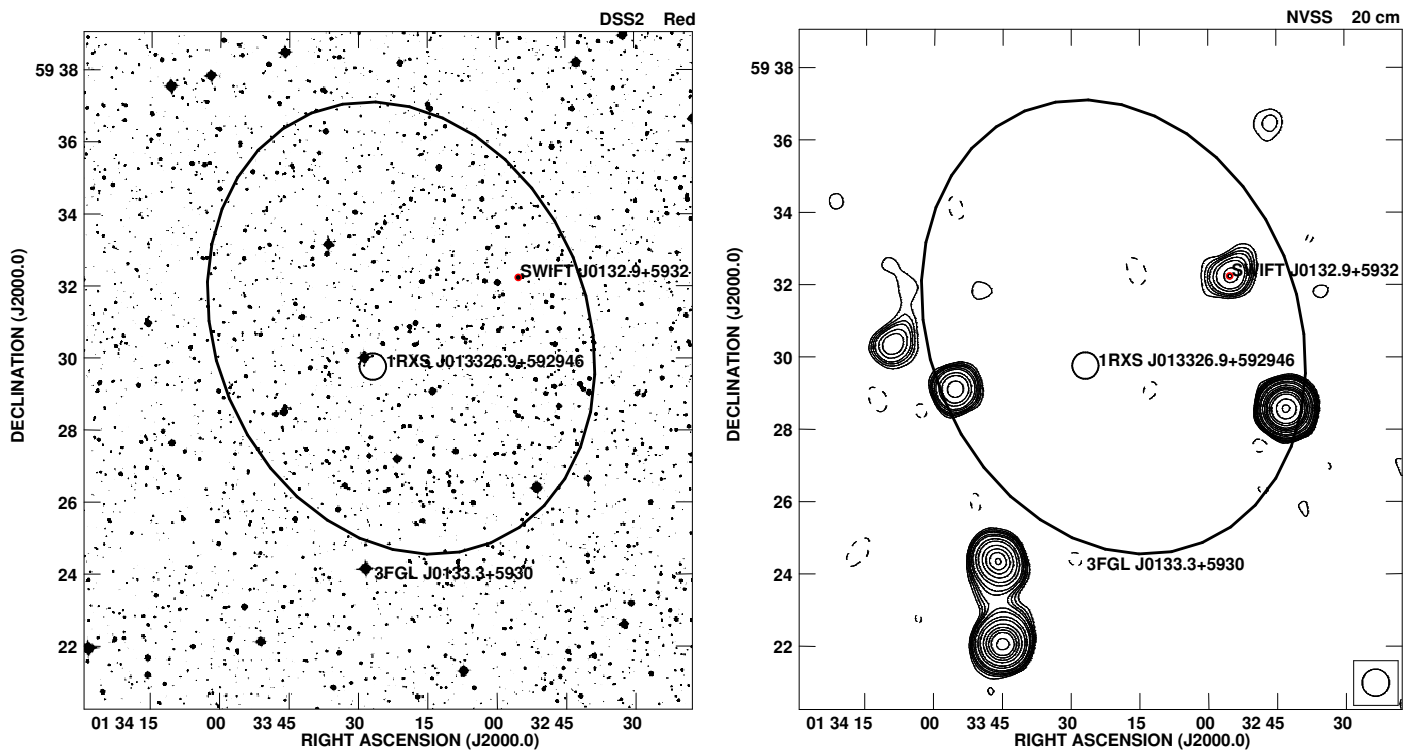}
      \caption{
 {\bf Left.} The 95\% confidence ellipse of  the gamma-ray source \fgl\  overplotted against the
      Second Digitized Sky Survey
      (red) distributed by the Space Telescope Science Institute. The small black and the tiny red circles
     indicate the positions of the {\it ROSAT} and {\it Swift} X-ray sources inside the {\it Fermi} LAT ellipse, respectively.
     \tyc\ is the brightest star at the edge of the {\it ROSAT} circle.
     {\bf Right.} The same field as it appears in radio at the 20 cm wavelength according to the NVSS. The contour levels shown correspond
     to $-3$, 3, 4, 5, 6, 8, 10, 15, 20, 30, 40, 50, 60, 80, 100, 150, and 200 times the rms noise level of 0.4 mJy beam$^{-1}$.
     The lower right circle illustrates the NVSS restoring  beam with $45^{\prime\prime}$ FWHM.     
    }
         \label{ellipses}
   \end{figure*}

Unassociated gamma-ray sources provide one of the greatest challenges in observational astronomy, when scientists are
confronted with the problem of determining which type  of celestial object is responsible for the detected high or very high energy photons
from an often poorly ($\sim$ 0\grp 1-0\grp 5) defined direction. 
More than 30\% of the approximately three thousand sources of GeV photons in the third {\it Fermi} Large Area Telescope (LAT) source catalogue (3FGL, \citealt{2015ApJS..218...23A}) lack 
an associated counterpart at lower energies. A similar problem exists for high-energy sources in the TeV domain,  such as those
detected using the current generation of  
Cherenkov telescopes. About 20\% of the TeV sources in the current version
of the TeVCat catalogue\footnote{http://tevcat.uchicago.edu} also lack a lower energy counterpart. 

In the past, some unassociated gamma-ray sources near the galactic equator were shown to belong to the selected class of gamma-ray binaries. 
These stellar systems are remarkable la\-boratories for high-energy astrophysics because 
most of their power is radiated beyond 1 MeV \citep{2013A&ARv..21...64D, 2013APh....43..301P}. Objects such as \object{LS 5039} and  \object{LS I +61 303} 
 provide very representative examples of this new type of sources \citep{2000Sci...288.2340P, 2005Sci...309..746A, 2006Sci...312.1771A}, 
 where the optically visible companion is typically a luminous early-type star.
 
In this context, the present work builds on our previous efforts to contribute
 to enlarge the known population of binary systems with a significant gamma-ray component in their
spectral energy distribution (SED). The general outline of our approach has been described in \citet{2015Ap&SS.356..277M}, and is based on a joint  cross-identification
of 3FGL gamma-ray sources against different optical, X-ray and radio catalogues. 

Here,  we report our study of the unassociated  {\it Fermi}  source \fgl, which we intensively explored in our quest for  new gamma-ray binaries.
The  selection process  of this 3FGL target and the follow-up observational analysis of its
95\%  confidence ellipse is presented. Inside it,
two peculiar objects have been found. The first is the Be star \object{TYC 3683-985-1}
and the second a {\it Swift} X-ray emitter. Their observational properties and possible association with the gamma-ray source are
addressed in the following sections.

\section{Target selection}

Our attention focused on \fgl\ as a continuation of our work in \cite{2015Ap&SS.356..277M}, which was aimed at selecting early-type stars that show indications 
of non-thermal signatures. 
The main source of optical stellar information came from the catalogues collectively know as
Luminous Stars in the Northern Milky Way (LS) that was pioneered by \citet{1959LS....C01....0H}.
After cross-correlating the 3FGL source list with LS catalogues, the Be star
\tyc\ was found to be well inside the 95\% confidence ellipse of this {\it Fermi} LAT gamma-ray source. 
Also known as \object{LS I +59~79}, this optical star is a bright B1/2Vnne object ($V=10.7$) according to \cite{1974AJ.....79..597M}.
Remarkably, these authors have  flagged this star as a possible variable object more than 40 years ago.
Based on the adopted spectral type and the observed magnitudes,  we estimate an approximate colour 
excess $E(B-V) \simeq 0.4$ mag and a distance of about 2 kpc.
In addition, \tyc\ appears
to be consistent with the X-ray source \rosat\ in the ROSAT All-Sky Survey (RASS) faint source catalogue \citep{2000IAUC.7432R...1V}.
All these facts rendered our original suspicion conceivable that \tyc\ could be 
related to the {\it Fermi} emission and prompted an intense observational
follow-up aimed to seek  further evidence.

In the two panels of Fig. \ref{ellipses}, we provide a general picture  of the {\it Fermi}, optical, and {\it ROSAT}  sources mentioned so far.
Other sources detected by the  {\it Swift} X-ray observatory \citep{2013ApJS..207...28S} 
and the NRAO VLA Sky Survey (NVSS, \citealt{1998AJ....115.1693C}) are also included. We refer again to them
in the sections below.


\section{Observational follow-up}

We started our observational efforts by carrying out optical photometry with two main goals. First, to confirm the possible \tyc\ 
variability suggested by \citet{1974AJ.....79..597M}, and second, to check whether this  variability revealed some type of periodicity that
could be related to the orbital cycle of a binary star. 
The \fgl\ variability index amounts to 42.6 in the latest {\it Fermi} catalogue, implying that $\gamma$-ray variability was not
confidently detected during the first four years of the mission.
In parallel, we also inspected the different archives of radio, X-ray, and gamma-ray
observatories to better assess the possible low-energy counterpart of \fgl\ from a multi-wavelength point of view. 


\subsection{Optical photometry}

A campaign of 45 nights of CCD photometric observations was conducted starting on 18 September 2015 and ending on 2 March 2016 using the 41 cm 
University of Ja\'en Telescope (UJT), equipped with a set of $UBVR_cI_c$ Johnson-Cousins filters. Despite being located inside an urban area,
this equipment enables
differential photometry measurements  with typical 0.01--0.02 mag accuracy for objects as bright as \tyc. We refer to 
\citet{2016A&A...586A..58M} for a more detailed description of the UJT instrumental setup and observation procedure. 
The absolute magnitudes of comparison stars in the field
were derived by performing absolute photometry  based on \citet{1992AJ....104..340L} standards on selected clear nights.

\begin{figure}
   \centering
   \includegraphics[angle=0,width=9.0cm]{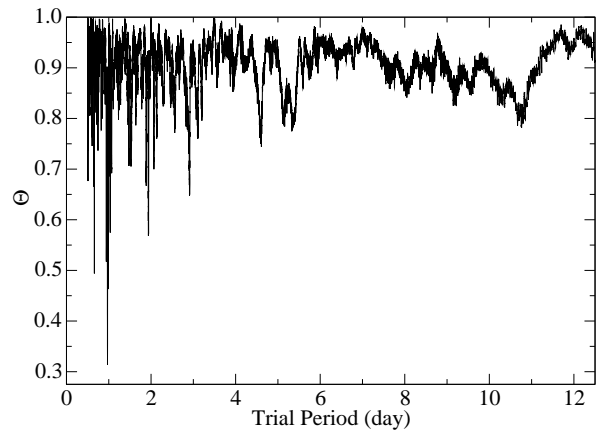}
      \caption{PDM periodogram of the $V$-band observations of \tyc\  where the statistic $\Theta$ displays
      a very deep minimum for a trial period of 0.97 d, which
      we interpret as half the true orbital period of the system.  
    }
         \label{pdm}
   \end{figure}

\begin{figure}
   \centering
   \includegraphics[angle=0,width=10.0cm]{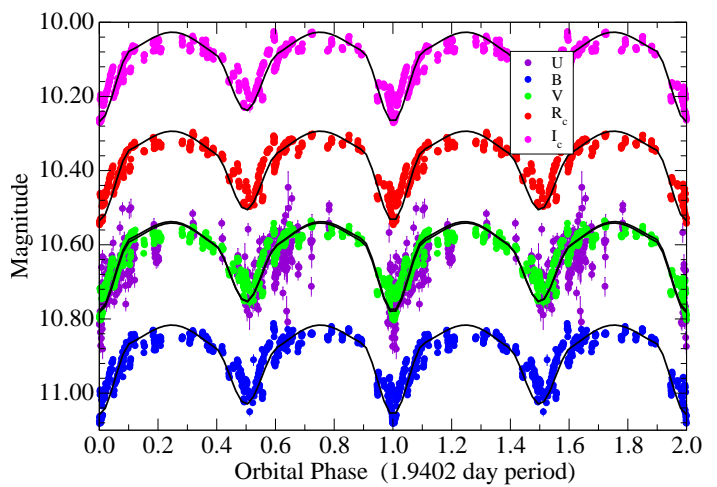}
      \caption{$UBVR_cI_c$ light curves of \tyc\ as observed with the UJT and folded using the orbital period value 
      of $1.9402$ d reported in this work.
      The continuous lines correspond to the synthetic light curves generated using the PHOEBE software packages with the physical
      parameters listed in Table \ref{phoebe}.  HJD 2457378.306 has been adopted as phase origin.
      All points are plotted twice for easier display. 
    }
         \label{ubvri}
   \end{figure}

During the first months, the UJT observations were performed on a daily basis and variability with a $\sim 0.1$ mag amplitude in all filters was discovered
with an apparent period of about 32 d. After January 2016, we searched for short-term variability by monitoring \tyc\ over intervals of several hours per night.
To our surprise, clear brightness changes were also detected at the $\sim 0.1$ mag level on timescales of hours. The acquisition of further hours of photometry
on nearly consecutive nights enabled us to perform a more accurate period search. 
The resulting periodogram  using the the phase dispersion minimization (PDM) method by \citet{1978ApJ...224..953S} is presented in Fig. \ref{pdm}.
This PDM plot clearly shows a deep minimum at $0.9701 \pm 0.0003$ d and higher harmonics. 
At this point, the situation became clearer as we realized that the 32 d photometric
modulation initially obtained from the 2015 data alone was, in fact, the beating period between the PDM minimum associated period and the Earth rotation. 
Although not shown in Fig. \ref{pdm}, this beating period is also recovered by the PDM technique.
The folded light curves were very reminiscent of an eclipsing binary system, where the true orbital period is twice the PDM result. Modelling of the light curves
to be discussed later strongly supports this interpretation. Therefore, the \tyc\ orbital period is established  to be
$1.9402 \pm 0.0006$ d. This value is also confirmed when other periodogram methods are used, such as 
the CLEAN and the Lomb-Scargle algorithms \citep{1987AJ.....93..968R, 1982ApJ...263..835S}.
Phase zero is set at HJD 2457378.306, which corresponds to the time of the first deepest minimum observed.
The whole UJT photometric data set is presented in Fig. \ref{ubvri} plotted as a function of orbital phase.



   \begin{figure}
   \centering
  \includegraphics[angle=0,width=9.0cm]{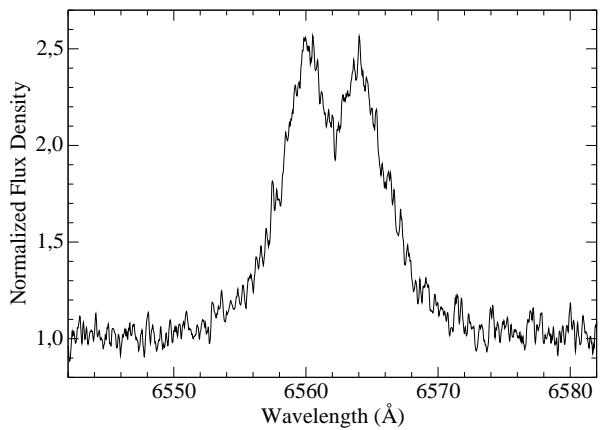}
      \caption{Optical spectrum of \tyc\ in the H$\alpha$ region as observed with the NOT telescope and the FIES instrument.
    }
         \label{halfa}
   \end{figure}

   \subsection{Optical spectroscopy}
   
 \tyc\ was observed with the Nordic Optical Telescope (NOT) on 3 February 2016.
 The FIES cross-dispersed high-resolution echelle spectrograph was used.
 The acquired frames were reduced and calibrated using the standard procedures embedded in the FIEStool pipeline 
 for automated data reduction. Because of unfortunate bad  weather conditions, only the red part of the spectrum remained barely acceptable for science purposes. In Fig. \ref{halfa} we display the H$\alpha$ emission line of \tyc, where a clear double-peak profile is evident. 
 By deblending the emission with two Voigt profile components\footnote{The Voigt profile results from the convolution of Lorentz and Gaussian profiles.},
 we estimate a peak-to-peak separation of $\Delta V \simeq 200$ km s$^{-1}$ and a  total equivalent width  $EW \simeq -16$ \AA.
 
 Another NOT spectrum was obtained on 22 September 2016  under better
 observing conditions with similar H$\alpha$ parameters. In addition, we were able to
 detect the He I lines at 4026, 4143 and 4387 \AA. The full-width at half-maximum (FWHM) of these absorption features is correlated with the 
 projected rotational velocity according to  \citet{1999A&AS..137..147S}. Using their expressions, we derive
 $v\sin{i} = 290 \pm 20$ km s$^{-1}$.

   
   


   


\subsection{Re-analysis of \tyc\  in archival X-ray data}

\begin{figure}
   \centering
   \includegraphics[angle=270,width=9.0cm]{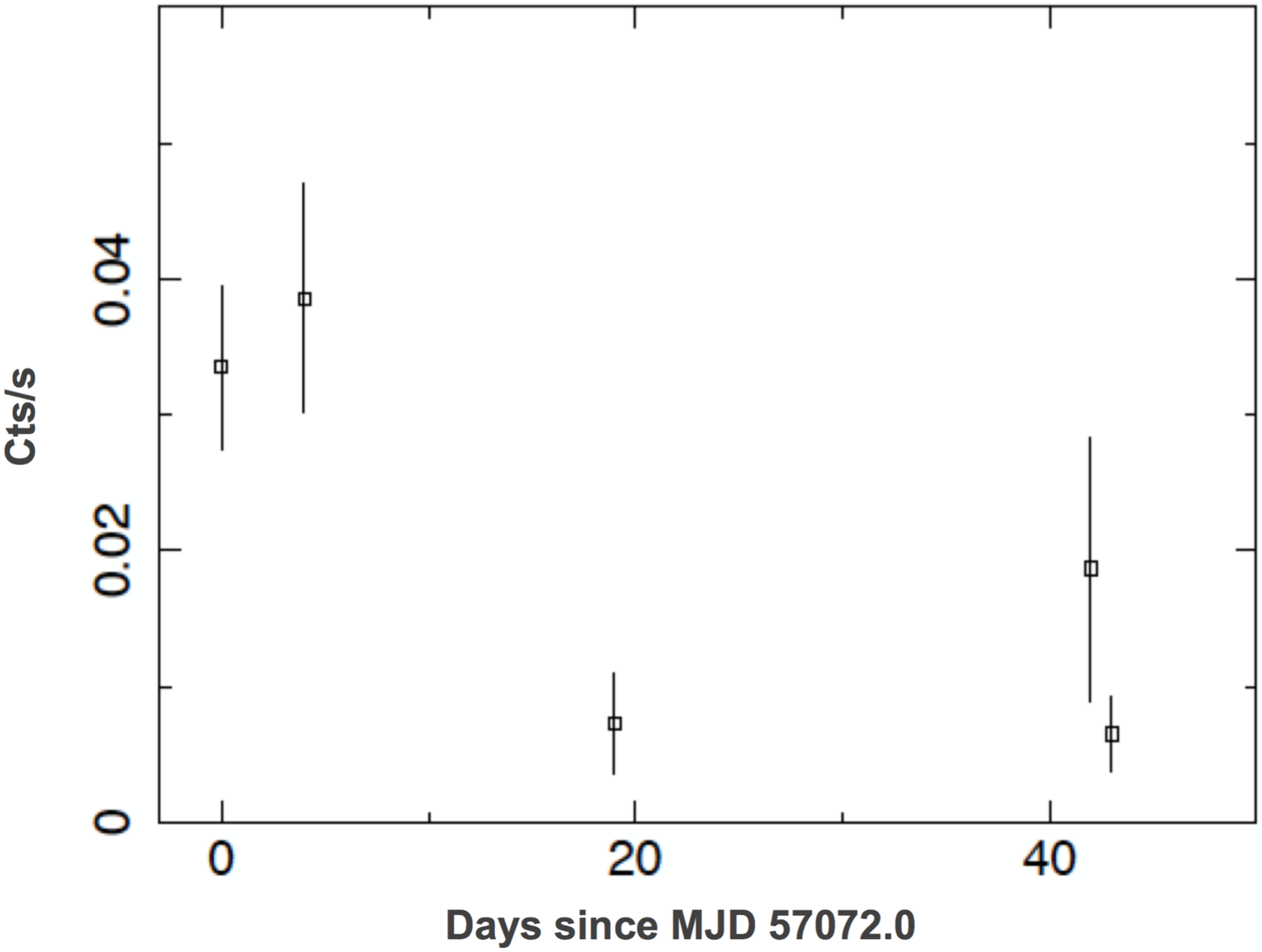}
      \caption{Light curve  of the X-ray source \sw\ inside the \fgl\ ellipse showing a clearly variable
      behaviour on timescales of weeks in the 0.2--10 keV energy range.
    }
         \label{xrayvar}
   \end{figure}

The software packages HEAsoft/FTOOLS and Xselect were used here for X-ray data reduction. 
We downloaded the original RASS event file corresponding to the \rosat\ detection
in the  0.1--2.4 keV energy range. Re-analysis of these data resulted in a very marginal detection, suggesting
that this X-ray source could  be spurious.  
Based on the star colour excess, the equivalent hydrogen column density is estimated to be $N_H \simeq 2.3 \times 10^{21}$ cm$^{-2}$
using standard relationships \citep{1978ApJ...224..132B}.
Assuming a typical photon index $\Gamma=2$, the source count rate of  about 0.038 counts s$^{-1}$ would translate into
an X-ray luminosity of approximately $9\times 10^{32}$ erg s$^{-1}$ in the ROSAT energy band.
From Table 2 of \citet{1996A&AS..118..481B}, we estimate the typical ROSAT luminosity of isolated Be stars as close to $10^{30.6 \pm 0.9}$ erg s$^{-1}$.
Therefore, the \rosat\ luminosity appears to be well above the high-end range of individual emission-line B stars.

At present, only the {\it Swift} X-ray observatory
has covered the \tyc\ field of view on different observing epochs, from 19 February to 3 April 2015, with a total on-source time of about 5 ks.
We also retrieved and re-analysed the corresponding {\it Swift} event files in the 0.2--10 keV energy range,  and were unable to recover the RASS source.
The count rate was lower than $0.003$ count s$^{-1}$ at the $3\sigma$ level.
The luminosity upper limit derived from {\it Swift} data would imply a ROSAT luminosity not above $8\times 10^{31}$ erg s$^{-1}$.
Therefore, the {\it ROSAT} X-ray source initially associated with \tyc\ is  either spurious or  variable by at least one order of magnitude.
In the second case, it would have been non-active during all the 2015 X-ray observing dates.

\subsection{Another X-ray source present in {\it Swift} data }

In parallel, the {\it Swift} telescope data clearly revealed a different X-ray source at the J2000.0 position
R.A. $01^h 32^m$55\rl 2 and DEC. $+59^{\circ} 32^{\prime}$14\pri 7 with a 90\% confidence error circle of 15 arcsecond.
Hereafter, we refer to it as \sw\ and this is one of the plotted objects in the panels of Fig. \ref{ellipses}. 
Its count rate amounts to  $0.020 \pm 0.003$ count s$^{-1}$ corrected for point-spread-function (PSF) effects, thus providing a $7.7\sigma$
detection. Its presence inside the {\it Fermi} ellipse has  been  pointed out  by \citet{2013ApJS..207...28S}.
The measured position is 4.75 arcmin away from
 \tyc\ where no X-rays were detected.

A simple power-law fit to the \sw\ spectrum provides a photon index $\Gamma = 2.5 \pm 0.5$, together with an equivalent hydrogen
column density of $N_H = (6.7 \pm 2.2) \times 10^{21}$ cm$^{-2}$. 
 This value translates into a visual extinction of $A_V = 3.6 \pm 1.2$ mag, which is equivalent to a colour excess of $E(B-V)=1.2 \pm 0.4$ mag.
Inspection of the \sw\ daily count rates indicates that this X-ray source is variable up to a factor of $\sim5$  on a timescale $\tau \sim 10$ d.
A plot of the observed light curve with PSF and vignetting corrections is presented in Fig. \ref{xrayvar}.

\begin{figure}
   \centering
   \includegraphics[angle=0,width=9.0cm]{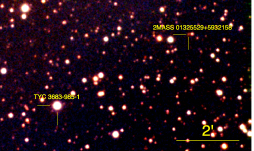}
      \caption{
   Combined  
       view of the \tyc\ field as observed with the UJT in the $VR_cI_c$ bands 
      (blue, green and red filters, respectively). The total integration time is about 1 h in each photometric band. The locations of \tyc\
      and the faint 2MASS object consistent with the \sw\ X-ray source are marked by yellow lines. The horizontal bar sets the angular scale. North is up
      and east is left.
          }
         \label{VRI}
   \end{figure}

To search for the optical counterpart of \sw, we combined a large number of the CCD  $BVR_cI_c$ frames acquired for the \tyc\ photometry, and performed astrometry
on the stacked image shown in Fig. \ref{VRI}. A faint counterpart is apparently consistent with the {\it Swift} X-ray source within a few arc-seconds. Using the
UJT, we could only detect it with the reddest filters ($B>19.5$, $V>18.2$, $R_c=17.0 \pm 0.1$, and $I_c=16.1 \pm 0.1$).
This detection is independently confirmed when inspecting the images from  the Two Micron All Sky Survey (2MASS) and its 
associated catalogue  \citep{2012yCat.2281....0C}. 
Here, the UJT counterpart is consistent with  \twomass. This infrared source  
also becomes brighter with increasing wavelength ($J= 15.425 \pm 0.082$, $H=14.703\pm 0.096$, and $K_s= 14.002 \pm 0.079$).
In the NED Extragalactic Database\footnote{\tt https://ned.ipac.caltech.edu} this object  is catalogued as a galaxy based on its
apparent extension.
It is also relevant to note here that both the {\it Swift} and 2MASS objects are in good position agreement with the radio source
\nvss\ (see Fig. \ref{ellipses}), which was also marginally detected during the Westerbork Northern Sky Survey (WENSS, \citealt{1997A&AS..124..259R}).

   \begin{figure}
   \centering
   \includegraphics[angle=0,width=9.0cm]{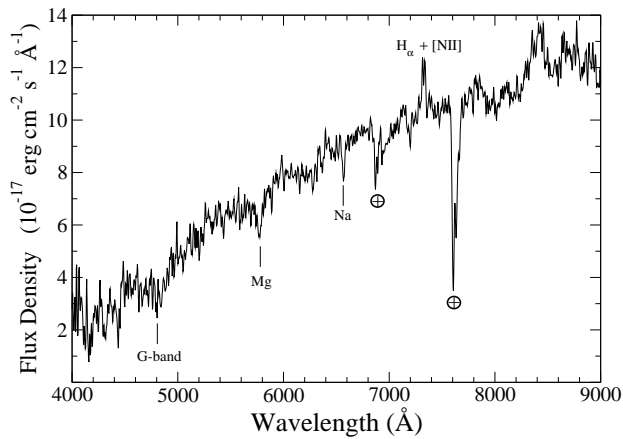}
      \caption{Optical spectrum of \twomass\ obtained with the NOT telescope and its ALFOS instrument on 21 September 2016.
      The most prominent absorption and emission features are identified. All of them are consistent with the redshift value stated in the text.}
         \label{not_alfosc}
   \end{figure}

To confirm the extragalactic nature of \twomass, we requested additional service time with the NOT telescope and its ALFOSC instrument.
The resulting spectrum of \twomass\ is displayed in Fig. \ref{not_alfosc}. Different absorption and emission features were detected and all of them are consistent
with a cosmological redshift of $z=0.1143 \pm 0.0002$. 
This would place the source at a luminosity distance of about 535 Mpc  assuming a Hubble constant $H_0 = 70$ km s$^{-1}$ Mpc$^{-1}$ in a flat universe
with $\Omega_{\Lambda}=0.28$.

\subsection{Re-analysis of {\it Fermi} archival data}

With the idea of searching for the 1.94 d period of TYC 3683-985-1 in {\it Fermi}-LAT data, we selected 
PASS 8 source class events with energies from 100 MeV to 500 GeV within a region of interest (ROI) of 
$10^{\circ}$ side length centred on the \fgl\ catalogue position. The LAT dataset 
spans from 2008-08-04 15:43:36 (UTC) to 2016-01-14 01:41:33 (UTC).
The analysis was performed with the {\it Fermi} ScienceTools software package v10r0p5. 
We used only events tagged as 'good' and excluded any event with zenith angle above 
$90^{\circ}$ to prevent Earth's limb contamination, as recommended by the LAT team. 
The reality of \fgl\ is confirmed with a test statistic value of $TS=27$. Its 
 gamma-ray spectrum has been obtained by means of a binned maximum likelihood analysis, yielding 
compatible results with those reported in the 3FGL catalogue \citep{2015ApJS..218...23A}.
Its power-law spectral index is $\Gamma = 2.6 \pm 0.3$, with a photon flux
of $(8\pm 3) \times 10^{-9}$ ph cm$^{-2}$ s$^{-1}$ in the 0.1--500 GeV range.

The fit results obtained for sources inside the ROI in the full dataset
have then been used to perform an unbinned analysis in time bins shorter than the orbital period. 
A binning interval of 0.9 d, about half the expected orbital period, was selected as a compromise between securing the largest photon counting and the Nyquist sampling rate.
As a result, the \tyc\ photometric period was not recovered from {\it Fermi} data in periodograms computed using both the PDM and CLEAN
methods. The most significant periodicities were consistently found at 1.81, 12.25 and 53.10 d. Their possible interpretation is at present beyond the scope
of this paper, and only a timescale order of magnitude is considered for discussion purposes.

\section{Discussion}

The observational data presented in previous sections have revealed two peculiar objects towards the unassociated gamma-ray source \fgl. 
Here, we discuss each of them in detail to establish their chances of association with the high-energy emission.

\subsection{Nature of \tyc}

Suggestive evidence about the binary nature of the star \tyc\  has been obtained from UJT observations given the existence of a well-determined
photometric period, which we associate with half the orbital cycle. 
To reinforce this interpretation, we carried out an attempt to model the $UBVR_cI_c$  light curves using the
PHOEBE software package \citep{2005ApJ...628..426P} based on the Wilson-Deviney code \citep{1971ApJ...166..605W}.
At this point, we noted that the plots of observed colours $U-B$, $B-V$, $V-R_c$, and $R_c-I_c$ (not shown here) remained
nearly constant throughout all orbital phases. This immediately suggested that the eclipsed photospheres had to be of very similar type. Therefore, 
the PHOEBE stellar parameters initially adopted were those of a twin pair of B-type main-sequence stars in agreement with the 
\cite{1974AJ.....79..597M} spectral classification. To keep the model as simple as possible, a circular orbit, no spots and the PHOEBE default values 
were adopted except where stated otherwise. After some iteration, it was soon possible to converge on a set of stellar parameters
that reasonably reproduce the overall appearance of the phased light curves in Fig. \ref{ubvri}. The main parameter values are given in Table \ref{phoebe}.

\begin{figure}
   \centering
   \includegraphics[angle=0,width=9.0cm]{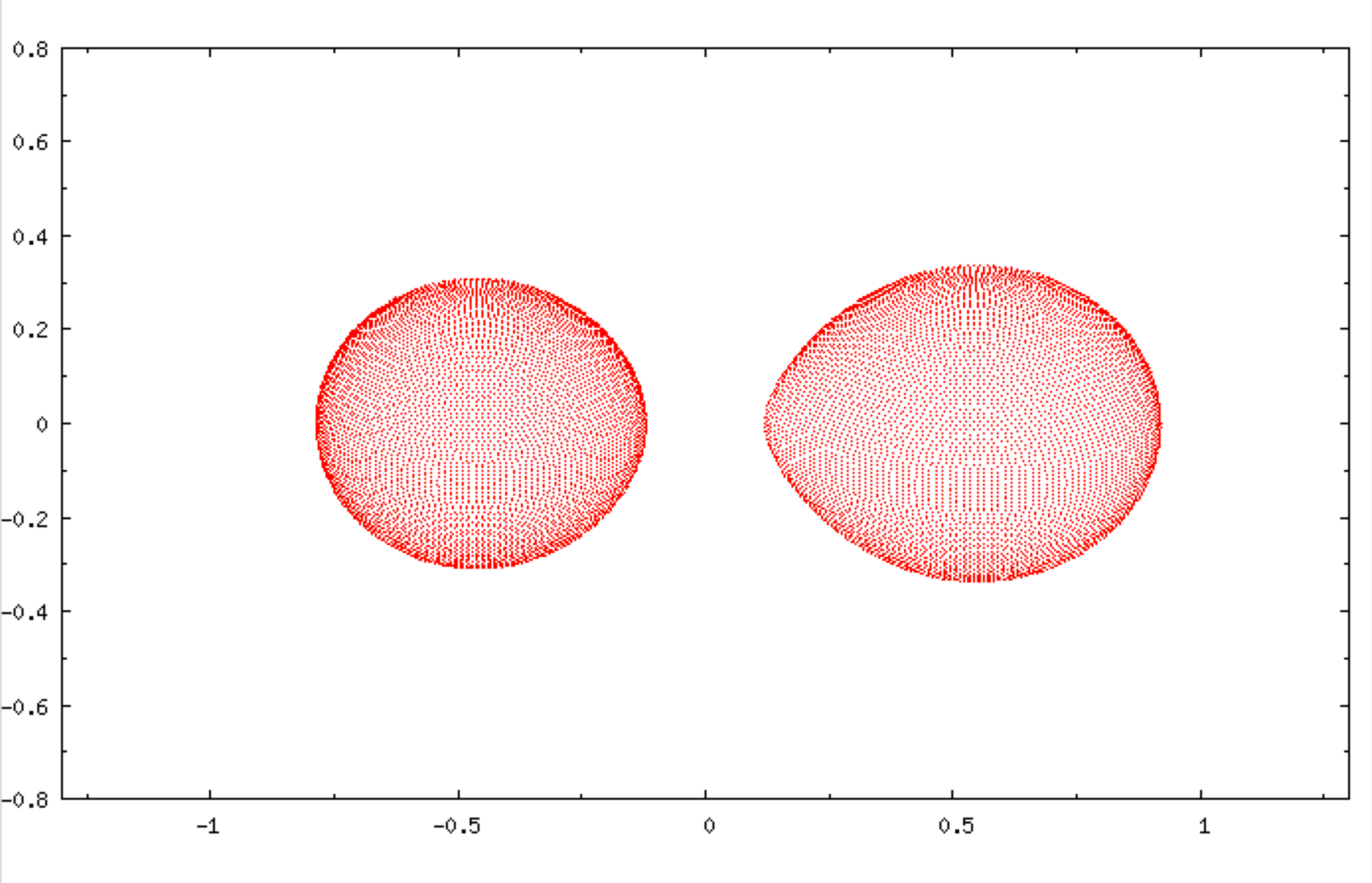}
      \caption{Shape of the \tyc\ stellar components resulting from the PHOEBE parameters 
      in Table \ref{phoebe}. 
      The primary and secondary stars are located at the left and right side of the plot,  respectively.
      The secondary star
      is practically filling its Roche lobe. Axes labels are given in units of the semimajor axis. 
    }
         \label{roche}
   \end{figure}

\begin{figure}
   \centering
  \includegraphics[angle=0,width=10.2cm]{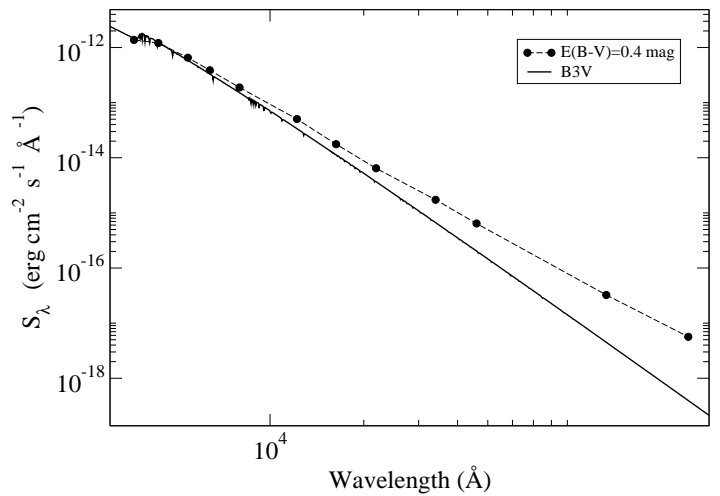}
      \caption{Dereddened SED of \tyc\ from optical to infrared wavelengths (dashed line) built from the UJT magnitudes in the maximum
      and far-infrared data from the WISE bands \citep{2010AJ....140.1868W}.
      The thick line has been plotted for comparison purposes. It corresponds to a Kurucz 
      model for an isolated B3V star normalised to the $B$-band emission level where the free-free excess from the circumstellar disc is almost negligible.
    }
         \label{sed1}
   \end{figure}

The PHOEBE fit overplotted in Fig. \ref{ubvri} provides support to the interpretation of \tyc\ as an eclipsing binary consisting of a pair
of similar, early-type, B-stars. However, it must be regarded as very preliminary since radial velocities from optical spectroscopy are
not yet available, and this is a key information to better constrain the mass ratio parameter. Nevertheless, the fit is highly indicative that the secondary
star is almost filling its Roche lobe, as illustrated by the plot of stellar shapes in Fig. \ref{roche}. 
We therefore interpret  a semi-detached binary system. 
Matter flowing across the inner
Lagrangian point ($L_1$) will form a circumstellar or circumbinary disc.
Our two spectroscopic observations
of \tyc\ using the NOT telescope, showing both double-peaked H$\alpha$ emission (see Fig. \ref{halfa}), agree well with this
interpretation and confirm the Be nature of the system. In this case, free-free emission from the Be disc, mostly in ionised form, is expected
to be responsible for an emission excess at long infrared wavelengths. This is well
visible in the SED of  \tyc\ plotted in Fig. \ref{sed1} as compared with  the expected photospheric continuum
 from an early B-type main-sequence Kurucz model\footnote{{\tt http://www.stsci.edu/hst/observatory/crds/k93models.\\ html}}.
 
 Assuming a Keplerian rotationally dominated profile, the  H$\alpha$ double-peak separation $\Delta V$ can provide a measure of the Be disc outer radius  
 given by \cite{1972ApJ...171..549H}:
 \begin{equation}
 R_{\rm disc} = R_1 \left[   \frac{2 v\sin{i}}{\Delta V}  \right]^2,
 \end{equation}
 where $R_1$ is the radius of the primary star and $v\sin{i}$ and $\Delta V$ parameters have been estimated from optical spectroscopy.
 Adopting  $R_1 = 4.2$ $R_{\odot}$ from our PHOEBE fit, the 
 expected circumstellar disc radius is $R_{\rm disc} \simeq 35$ $R_{\odot}$.
 This value exceeds the semi-major
 axis given in Table \ref{phoebe}, thus suggesting that  this may be a circumbinary disc in \tyc.
  
Finally, the brightness level in the light curves in Fig. \ref{ubvri} does not have a strict repeatability from orbit to orbit, with 0.01--0.05 mag minor differences.
We tentatively attribute this to physical effects not taken into account
in the binary model, such as Be disc variability, a time-variable hot spot, or intrinsic variability of one of the stellar components.

\begin{table}
\caption{Values of the main PHOEBE parameters for \tyc}          
\label{phoebe}      
\centering                        
\begin{tabular}{l l c }     
\hline\hline               
Parameter  & Value   &   Comments \\
 \hline
Eccentricity         & $e=0.000$ & fixed \\
Semimajor axis           & $a=13.1 \pm 0.4$ $R_{\odot}$  &     \\
Mass ratio   &  $\frac{M_2}{M_1}=0.845 \pm 0.004$  &   \\
Inclination         &  $i= 65.9 \pm 0.1^{\circ}$   &  \\
Primary effective  &  $T_1 = 19000 \pm 500$ K  &    \\
temperature & & \\
Secondary effective           &    $T_2 = 21000 \pm 600$ K              &           \\
temperature &  & \\
Primary star  & $\Omega_1=4.1 \pm 0.1$ & $(^a)$  \\
surface potential &  & \\
Secondary star & $\Omega_2=3.55 \pm 0.05$ &   $(^a)$ \\
surface potential &  & \\
\hline
Potential value at the  & $\Omega_{L_1}= 3.49$ &   computed  \\
inner Lagrangian point &  &     \\
Potential value at the & $\Omega_{L_2}= 3.02$ & computed  \\
second Lagrangian point &   &  \\
\hline
\end{tabular}
\tablefoottext{a}{Kopal modified potential as defined in \citet{kopal1959close}.}
\end{table}


\subsection{Nature of \sw\ / \twomass}
  
\begin{figure}
   \centering
  \includegraphics[angle=0,width=9.5cm]{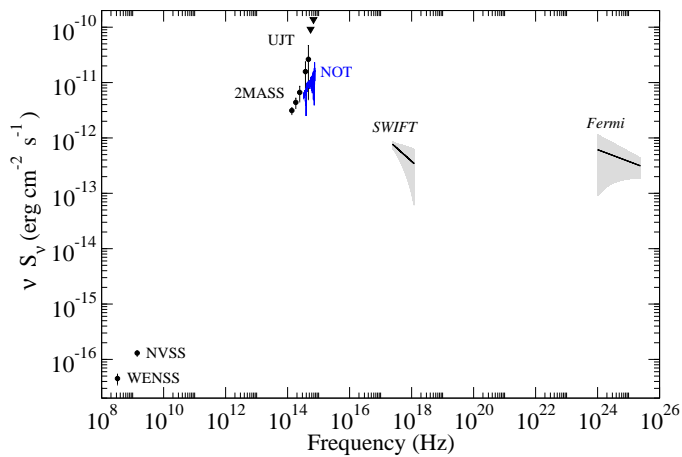}
      \caption{De-reddened SED of the X-ray source \sw\ also consistent with \twomass\ and \nvss, which has been proposed
      as a possible counterpart of \fgl. Data shown include  measurements  at radio (WENSS and NVSS),
      near-infrared (2MASS), optical (UJT and NOT), X-ray ({\it Swift}) and gamma-ray ({\it Fermi}) frequencies.
      Triangle symbols represent $3\sigma$ upper limits
      for the $B$- and $V$-band non-detections with UJT. 
        }
         \label{sed2}
   \end{figure}

 The other peculiar object inside the {\it Fermi} confidence ellipse is the infrared source \twomass, found to be coincident
 with the variable X-ray source \sw\ and also with \nvss\ as a likely radio counterpart.
 The 2-10 keV luminosity of the X-ray source is $1.7\times 10^{43}$ erg s$^{-1}$.
 Combining the X-ray properties, the optical spectrum and the multi-wavelength SED (see Fig. \ref{sed2}), an active galactic nucleus (AGN) 
 of BL Lac type can offer a natural explanation for both X-ray and gamma-ray emission detected by {\it Swift} and {\it Fermi}, respectively. 
 The SED would be a sum of blazar emission and stellar light of the host galaxy dominating in the near-infrared to optical band. 
 The X-ray emission can be understood as the high-frequency side of a synchrotron hump and its steep slope ($\Gamma\simeq 2.5$) 
 matches it well while the gamma-ray emission would be the inverse-Compton component with a comparable luminosity. 
 This and the lack of broad-line region (BLR) in the Fig. \ref{not_alfosc} optical spectrum agree with expectations from a BL Lac with a radiatively inefficient disc 
 \citep[see][]{galaxies4040036}.
 The host galaxy has a luminosity a few times of $\sim 10^{44}$ erg s$^{-1}$, suggesting  
 $\sim 10^9 M_{\odot}$ for the black hole mass of the blazar, which is consistent with the low radiative efficiency mentioned above. 
 It probably has moderate star formation with a star formation rate $\sim 1 M_{\odot}$ yr $^{-1}$ estimated from the 
 H$_{\alpha }$ luminosity \citep{1994ApJ...435...22K}. 
 
 On the other hand, if the {\it Swift} source is associated with a normal AGN, that is one of
 Seyfert type, which would not produce detectable gamma-ray emission, then some problems may arise. 
 The lack of strong absorption, except for the moderate Galactic absorption, in the {\it Swift} XRT spectrum indicates little obscuration 
 in the line of sight towards the nucleus. It contradicts the absence of BLR and strong nuclear emission in the optical spectrum. 
 The steep X-ray slope is also uncommon for a normal Seyfert galaxy, although the error on the slope is large.

\subsection{Quest for a gamma-ray candidate: binary or AGN?}

Two possible objects have emerged as low-energy counterpart candidates to the unassociated source \fgl.
 \sw\ could provide a classical blazar-like scenario,  but with the available data we cannot rule out a normal AGN, which would be unable to account
 for the {\it Fermi} photon flux at the measured redshift. This leaves the possibility that this might be a stellar source of gamma rays, given the co-location
 of \fgl\ with the Be binary \tyc. However, this system appears to consist of
two non-degenerated stars revolving in an eclipsing orbit, which at first glance makes it challenging to attribute it to the origin
of  high-energy emission. It would not correspond to the normal scenario of a gamma-ray binary (compact companion + non-degenerate star).  

Caution is nevertheless advisable here since \tyc\  could turn out to be unusual in some aspect, such as having a highly magnetised 
early-type component where acceleration of particles
could occur in shocks created by strong magnetic activity and powerful magnetic reconnection events.
 A similar mechanism has been proposed to account for collective gamma-ray
emission in stellar magnetospheres of T Tauri stars \citep{2011ApJ...738..115D}. We point out here that magnetic fields higher than those of T Tauri stars
(up to $\sim10$ kG) have been reported for some OB stars, and  they are sometimes  even highly variable on hourly timescales \citep{2015A&A...578L...3H}.

Alternatively,  the eclipsing binary nature of \tyc\  may not have
been correctly interpreted, and it might be a compact companion orbiting a single Be star.
Gamma-ray emission would then be naturally expected. This was initially considered
because the observed light curves bear some resemblance to ellipsoidal modulation that would
result from a nearly Roche-lobe filling star. However, an ellipsoidal modulation tends to have a more sinusoidal aspect and smaller
amplitude (few 0.01 mag),  and this is not the case
of the light curves in our Fig. \ref{ubvri}.

Other radio sources inside the \fgl\ ellipse were also considered at some point, but none of them
displays X-ray emission, and no evidence of an optically variable counterpart is available.

\section{Conclusions}

A multi-wavelength analysis of peculiar objects inside the 
confidence ellipse of the unassociated gamma-ray source \fgl\
has been carried out. As a result, only one pair of objects have attracted our attention as possible low-energy counterparts to the
{\it Fermi} emission. 

One of them is the infrared source \twomass\ that is also detected in X-rays as \sw. We found this object to be an
AGN at  $z=0.1143 \pm 0.0002$. The observed SED agrees with expectations from a blazar and this could
provide an easy interpretation of the high-energy emission. However, other
less collimated AGN interpretations unable to account for the gamma-ray luminosity  (e.g. a Seyfert)  can also hold. 

The alternative counterpart candidate is the Be star \tyc. After an extensive campaign of photometric observations,
we have discovered apparent eclipses with a clear period of $1.9402 \pm 0.0006$ d that we interpret as the orbital cycle.
This period has not been detected yet in the {\it Fermi} emission. However, the gamma-ray periodicity analysis was not conclusive enough
because, with the current data, the time binning had to be  extremely close to the Nyquist sampling limit.
Modelling of the \tyc\ optical light curves is consistent with a semi-deteached Be binary system  whose two components are 
 early-type B stars, one of them filling its Roche lobe. In addition,
the observed double-peaked H$\alpha$ profile suggests that the Be disc is circumbinary.
Different physical mechanisms for high-energy photon production in a Be stellar environment have been tentatively proposed.
In any case, the non-degenerate nature of the two components remains to be confirmed since the possibility of one
of them being a compact object cannot be strictly ruled out from photometric data only. 

Solving the final nature of \fgl\ will require additional follow-up observations.
VLBI imaging of \sw\ is likely to settle the question whether it is a
 highly beamed source. Similarly, radial velocity observations of \tyc\ will enable a revision
of the component masses, which is necessary to finally assess whether the system is capable of gamma-ray production.

\begin{acknowledgements}
    This work was supported by grants AYA2013-47447-C3-1-P,  AYA2013-47447-C3-2-P, 
    AYA2013-47447-C3-3-P, and FPA2015-69210-C6-2-R from the Spanish Ministerio de Econom\'{\i}a y Competitividad (MINECO),
and by the Consejer\'{\i}a de Econom\'{\i}a, Innovaci\'on, Ciencia y Empleo of Junta de Andaluc\'{\i}a
under excellence grant FQM-1343 and
research group FQM-322, as well as FEDER funds.
JMP, KI, and MR acknowledge support from MINECO under grant MDM-2014-0369 of ICCUB (Unidad de Excelencia 'Mar\'{\i}a de Maeztu'), and the Catalan DEC grant 2014 SGR 86. 
Part of this paper is based on observations made with the Nordic Optical Telescope, operated by the Nordic Optical Telescope Scientific Association at the Observatorio del Roque de los Muchachos, La Palma, Spain, of the Instituto de Astrof\'{\i}sica de Canarias.
This research has made use of the SIMBAD database,
operated at CDS, Strasbourg, France. 
This publication makes use of data products from the Two Micron All Sky Survey, which is a joint project of the University of Massachusetts and the Infrared Processing and Analysis Center/California Institute of Technology, funded by the National Aeronautics and Space Administration and the National Science Foundation.
We have made use of the WSRT on the Web Archive. The Westerbork Synthesis Radio Telescope is operated by the Netherlands Institute for Radio Astronomy ASTRON, with support of NWO.
We finally thank Valent\'{\i} Bosch-Ramon (ICCUB) for fruitful discussions on SED interpretation. 
\end{acknowledgements}

%
%


\bibliographystyle{aa} 
\bibliography{references.bib} 

%

\end{document}